\documentclass[iop]{emulateapj}

\usepackage{graphicx}
\usepackage{graphics}
\usepackage{txfonts}
\usepackage{epsfig}
\usepackage{natbib}
\usepackage{times}
\usepackage{amssymb}
\usepackage{color}
\usepackage{longtable}

\newcommand\ergcms{erg\,cm$^{-2}$\,s$^{-1}$}

\newcommand\ergs{erg\,s$^{-1}$}

\newcommand\cmsq{cm$^{-2}$}
\newcommand\integ{{\it{INTEGRAL}}}
\newcommand\swift{{\it{Swift}}}
\newcommand\maxi{{\it{MAXI}}}
\newcommand\xmm{{\it{XMM-Newton}}}
\newcommand\chan{{\it{Chandra}}}

\newcommand\nustar{{\it{NuSTAR}}}

\newcommand\nh{$N_\mathrm{H}$}

\shorttitle{\emph{NuSTAR} survey of the Norma Arm}
\shortauthors{Bodaghee et al.}

\begin{document}

\title{Initial results from \emph{NuSTAR} observations of the Norma Arm}

\author{Arash Bodaghee\altaffilmark{1,2}}
\author{John A. Tomsick\altaffilmark{1}}
\author{Roman Krivonos\altaffilmark{1}}
\author{Daniel Stern\altaffilmark{3}}
\author{Franz E. Bauer\altaffilmark{4,5,6}}
\author{Francesca M. Fornasini\altaffilmark{7}}
\author{Nicolas Barri\`{e}re\altaffilmark{1}}
\author{Steven E. Boggs\altaffilmark{1}}
\author{Finn E. Christensen\altaffilmark{8}}
\author{William W. Craig\altaffilmark{1,9}}
\author{Eric V. Gotthelf\altaffilmark{10}}
\author{Charles J. Hailey\altaffilmark{10}}
\author{Fiona A. Harrison\altaffilmark{11}}
\author{Jaesub Hong\altaffilmark{12}}
\author{Kaya Mori\altaffilmark{10}}
\author{William W. Zhang\altaffilmark{13}}

\altaffiltext{1}{Space Sciences Laboratory, 7 Gauss Way, University of California, Berkeley, CA 94720, USA}
\altaffiltext{2}{Georgia College \& State University, CBX 82, Milledgeville, GA 31061, USA}
\altaffiltext{3}{Jet Propulsion Laboratory, California Institute of Technology, Pasadena, CA 91109, USA}
\altaffiltext{4}{Instituto de Astrof\'{i}sica, Facultad de F\'{i}sica, Pontifica Universidad Cat\'{o}lica de Chile, 306, Santiago 22, Chile}
\altaffiltext{5}{Millennium Institute of Astrophysics, Vicu\~{n}a Mackenna 4860, 7820436 Macul, Santiago, Chile} 
\altaffiltext{6}{Space Science Institute, 4750 Walnut Street, Suite 205, Boulder, CO 80301, USA}
\altaffiltext{7}{Astronomy Department, University of California, Berkeley, CA 94720, USA}
\altaffiltext{8}{DTU Space, National Space Institute, Technical University of Denmark, Elektrovej 327, DK-2800 Lyngby, Denmark}
\altaffiltext{9}{Lawrence Livermore National Laboratory, Livermore, CA 94550, USA}
\altaffiltext{10}{Columbia Astrophysics Laboratory, Columbia University, New York, NY 10027, USA}
\altaffiltext{11}{Cahill Center for Astronomy and Astrophysics, California Institute of Technology, Pasadena, CA 91125, USA}
\altaffiltext{12}{Department of Astronomy, Harvard University, 60 Garden Street, Cambridge, MA 02138, USA}
\altaffiltext{13}{NASA Goddard Space Flight Center, Greenbelt, MD 20771, USA}

\begin{abstract}
Results are presented for an initial survey of the Norma Arm gathered with the focusing hard X-ray telescope \nustar. The survey covers 0.2\,deg$^{2}$ of sky area in the 3--79\,keV range with a minimum and maximum raw depth of 15\,ks and 135\,ks, respectively. Besides a bright black-hole X-ray binary in outburst (\object{4U 1630$-$47}) and a new X-ray transient (\object{NuSTAR J163433$-$473841}), \nustar\ locates three sources from the \chan\ survey of this region whose spectra are extended above 10\,keV for the first time: \object{CXOU J163329.5$-$473332}, \object{CXOU J163350.9$-$474638}, and \object{CXOU J163355.1$-$473804}. Imaging, timing, and spectral data from a broad X-ray range (0.3--79\,keV) are analyzed and interpreted with the aim of classifying these objects. \object{CXOU J163329.5$-$473332} is either a cataclysmic variable or a faint low-mass X-ray binary. \object{CXOU J163350.9$-$474638} varies in intensity on year-long timescales, and with no multi-wavelength counterpart, it could be a distant X-ray binary or possibly a magnetar. \object{CXOU J163355.1$-$473804} features a helium-like iron line at 6.7\,keV and is classified as a nearby cataclysmic variable. Additional surveys are planned for the Norma Arm and Galactic Center, and those \nustar\ observations will benefit from the lessons learned during this pilot study. \\
\end{abstract}



\vspace{10mm}
\section{Introduction}
\setcounter{footnote}{0}

The Norma Arm is among the most active regions of massive star formation in the Milky Way \citep{bro00}. It is not surprising that this region is also densely populated with the evolutionary byproducts of massive stars, neutron stars (NSs) and black holes (BHs). Many of these compact objects belong to binary systems and accrete matter from a normal stellar companion. These systems are called X-ray binaries (XRBs) and they represent laboratories for studying the physics of matter subjected to extreme gravitational and electromagnetic potentials. Their numbers can be used to constrain rates of massive star formation \citep[e.g.,][]{ant10}, while their spatial distributions are important for studies of stellar evolution \citep[e.g.,][]{bod12}. 

One advantage of surveying the Norma Arm is that it represents an intersection of molecular clouds, star-forming regions, and accreting compact objects, thereby providing X-ray source populations at various stages of evolution. These populations can then be compared with large populations residing in other active regions of the Galaxy such as the Galactic Center \citep{mun09} and Carina Arm \citep{tow11}. 

Thus, the Norma Arm has been the subject of recent observing campaigns seeking to uncover its X-ray populations. In the soft X-rays ($\lesssim$10\,keV), the \chan\ telescope discovered $\sim$1100 sources in a 1.3\,deg$^{2}$ section of this field. The largest source groups are cataclysmic variables (CVs), background active galactic nuclei (AGN), and stars (flaring, foreground, or massive), with other source types represented in smaller numbers (e.g., XRBs, young massive clusters, and supernova remnants: Fornasini et al., 2014, subm.). In the hard X-rays, \integ\ \citep[e.g.,][]{bir10,kri12} discovered a few dozen sources in the Norma Arm, almost all of which are XRBs.

With the advent of the hard X-ray focusing telescope \nustar\ \citep{har13}, it is now possible to map this region with unprecedented angular  ($18^{\prime\prime}$ full-width-half-maximum, 58$^{\prime\prime}$ half-power diameter) and spectral resolution (400\,eV) around 10\,keV. This paper presents results from a \nustar\ survey of a small section of the Norma Arm that took place in 2013 February. Section\,\ref{sec_obs} describes the analysis procedures employed on the \nustar\ data and on selected data from \chan, as well as some of the challenges inherent in X-ray observations of this field. In Section\,\ref{sec_res}, results from imaging, spectral, and timing analyses are presented for X-ray sources detected in the survey. Their implications on source classifications for these objects are discussed in Section\,\ref{sec_disc}.

\begin{deluxetable*}{ c c c c c c c }
\tablewidth{0pt}
\tabletypesize{\scriptsize}
\tablecaption{Journal of \emph{NuSTAR} observations of the Norma Arm.}
\tablehead{
\colhead{observation ID} & shorthand & \colhead{pointing R.A. (J2000)} & \colhead{pointing decl. (J2000)} 	& \colhead{position angle [deg]}	& \colhead{start date [UTC]} 	& \colhead{exposure time [s]} }
\startdata

40014001001			& 1 & 248.4829 & $-$47.7204 		& 160.15 			& 2013-02-24 01:46:07 	& 18407	\\

40014002001			& 2 & 248.3623	& $-$47.6444		& 160.15			& 2013-02-24 11:31:07 	& 19497	\\

40014003001			& 3 & 248.2407	& $-$47.5669		& 160.13			& 2013-02-21 20:31:07 	& 20846	\\

40014004001			& 4 & 248.5977	& $-$47.6374		& 160.12			& 2013-02-22 07:46:07 	& 19440	\\

40014005001			& 5 & 248.4775	& $-$47.5622 		& 160.13			& 2013-02-22 17:31:07 	& 21241	\\

40014006001			& 6 & 248.3529	& $-$47.4868		& 160.14			& 2013-02-23 04:46:07 	& 18959	\\

40014007001			& 7 & 248.7099	& $-$47.5554		& 160.14			& 2013-02-23 14:31:07 	& 22640	\\

40014008002			& 8 & 248.5845	& $-$47.4826		& 160.12			& 2013-02-20 23:31:07 	& 16573	\\

40014009001			& 9 & 248.4670	& $-$47.4038		& 160.12			& 2013-02-21 10:46:07 	& 14653
\enddata
\label{tab_log}
\end{deluxetable*}

\section{Observations \& Data Analysis}
\label{sec_obs}

\subsection{\emph{NuSTAR} data}

The \nustar\ data consist of nine pointings whose details are summarized in Table\,\ref{tab_log}. These nine pointings are comprised of two focal plane modules A and B (FPMA and FPMB) each having a field-of-view (FOV) of $13^{\prime} \times 13^{\prime}$. To increase sensitivity, adjacent pointings were tiled with significant overlap ($\sim$50\%) resulting in sky region covered by the survey of around 0.2\,deg$^{2}$ ($0.4^{\circ}\times0.4^{\circ}$), centered at (J2000.0) R.A. $=16^{\mathrm{h}} 33^{\mathrm{m}} 47^{\mathrm{s}}$ and decl. $= -47^{\circ} 32^{\prime} 14^{\prime\prime}$. In Galactic coordinates, this is $l=336.7776^{\circ}$ and $b=0.1825^{\circ}$.

Data analysis relied on HEASoft 6.14 and the \nustar\ Data Analysis Software (NuSTARDAS 1.2.0\footnote{http://heasarc.gsfc.nasa.gov/docs/nustar/analysis/nustardas\_swguide\_v1.5.pdf}) with the latest calibration database files (CALDB: 2013 August 30). Raw event lists from FPMA and FPMB were reprocessed using \texttt{nupipeline}\footnote{http://heasarc.gsfc.nasa.gov/docs/nustar/analysis} in five energy bands: 3--10\,keV, 3--79\,keV, 10--40\,keV, 10--79\,keV, and 40--79\,keV.

\subsubsection{Image cleaning}

Given the density of bright sources and the high level of diffuse background, the Norma Arm presents a number of unique challenges for \nustar. The first challenge is from the telescope mast which allows photons to land on the detector without having passed through the focusing optics. These are known as stray-light photons (a.k.a. 0-bounce photons) which originate from bright sources situated a few degrees outside the FOV of each module. Fortunately, these pixels are easily modeled and excluded by creating polygonal region files in \texttt{ds9} that correspond to the geometric patterns expected from stray light of known bright sources near the FOV. The main source of stray light for these Norma survey observations is \object{IGR J16320$-$4751} \citep{tom03,rod06}, a variable but persistent supergiant X-ray binary located between 0.2$^{\circ}$ and 0.5$^{\circ}$ outside the FOV.

The second, and more daunting, challenge was that \object{4U 1630$-$47}, a black-hole X-ray binary, was undergoing an outburst which means it was especially bright during our observations of this field \citep[$\sim$0.3\,Crab in 3--10\,keV:][see also \citet{kin14}]{bod12b}. When the source is outside the FOV, photons can still arrive on the detector modules without being properly focused. Such photons are called ghost-rays (a.k.a. 1-bounce photons), and their pattern is not completely understood \citep{kog11,har13}.

\begin{figure*}[!t]
\begin{center}
\includegraphics[width=\textwidth]{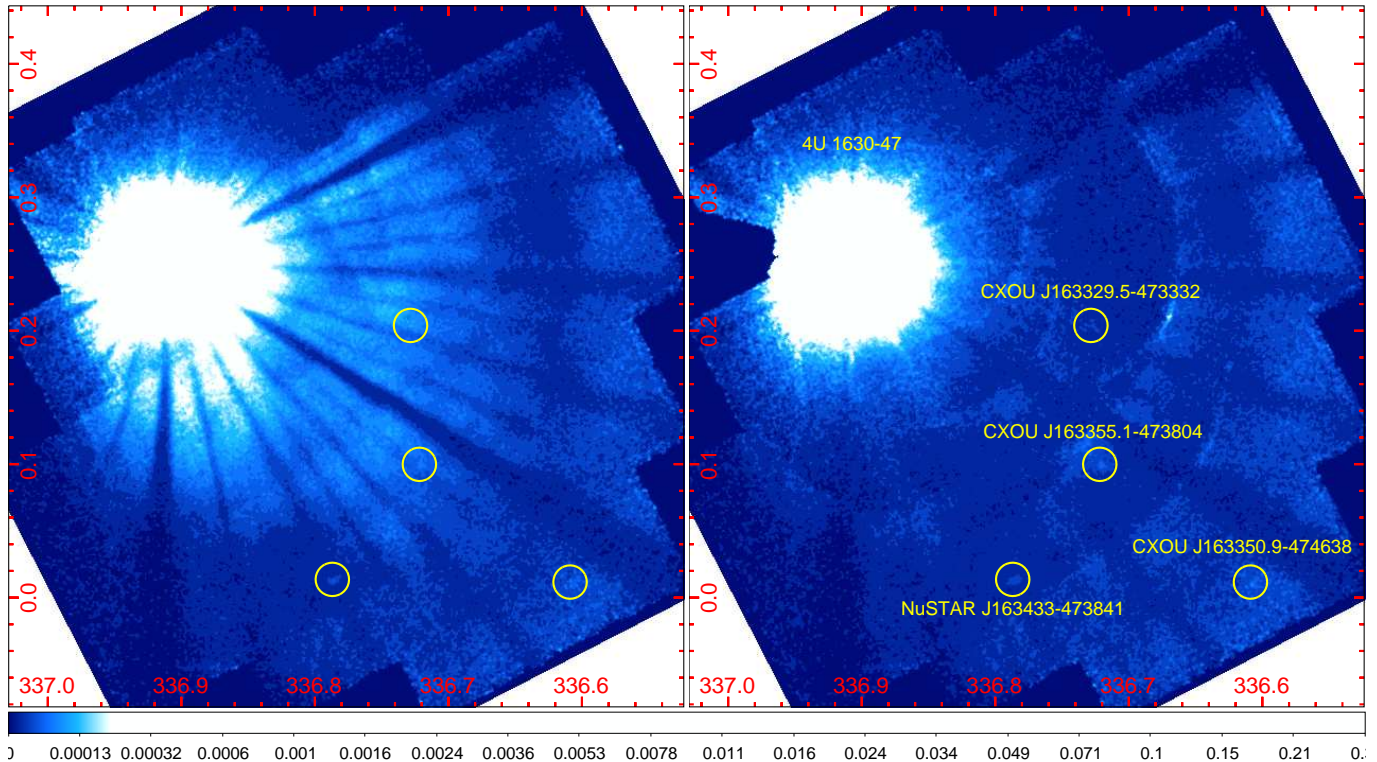}
\end{center}
\caption{Mosaic images (3--79\,keV) from the initial \nustar\ survey of the Norma Arm before (\emph{left}) and after (\emph{right}) the exclusion of pixels corresponding to stray light and ghost rays. Presented in Galactic coordinates, these exposure-normalized images (flux maps) combine 9 pointings with two focal-plane modules. The images are smoothed with a Gaussian kernel of $\sigma = 6^{\prime\prime}$ (each pixel is 2$^{\prime\prime}$ wide), and they are scaled logarithmically with an exaggerated contrast to aid visual identification of the detected sources (indicated with circles of radius $=45^{\prime\prime}$).}
\label{fig_mos}
\end{figure*}

In order to generate a mosaic image of the entire field where such effects could be minimized, we created new event lists (and exposure maps) in which we excluded regions with pixels contaminated by either stray light or ghost rays. This was done by visually examining the event lists of each observation (showing only those pixels, binned in blocks of 4, with more than 20 counts) and creating a polygonal region file in \texttt{ds9} that encompasses clusters of pixels (from both modules) on which ghost rays had fallen. By design, the regions were a few pixels wider than necessary to account for both the slightly different sky fields seen by each detector module and to account for the slight jitter due to the motion of the telescope mast. These cleaned event lists were used to generate an exposure-corrected mosaic image in the five energy bands listed above. Vignetting corrections were not applied to these mosaic images.

\begin{deluxetable*}{ c c c c c c }
\tablewidth{0pt}
\tabletypesize{\scriptsize}
\tablecaption{\emph{NuSTAR}-detected sources and their angular separation from likely \emph{Chandra} counterparts.}
\tablehead{
\colhead{name} & \colhead{R.A. (deg)}  & \colhead{decl. (deg)} & \colhead{90\% confidence radius} & \colhead{detection significance ($\sigma$)} & \colhead{offset}  }
\startdata

\object{CXOU~J163329.5$-$473332}	& 248.37254 & $-$47.55894 	& 7\farcs9 & 8.3 & 1\farcs9 	\\

\object{CXOU~J163350.9$-$474638}	& 248.46158 & $-$47.77642 	& 13\farcs0 & 15.0 & 4\farcs1 	\\

\object{CXOU~J163355.1$-$473804}	& 248.48046 & $-$47.63520	& 7\farcs0 & 8.7 & 2\farcs7 

\enddata
\tablecomments{Results for two other sources detected by \emph{NuSTAR}, \object{4U 1630$-$47} and \object{NuSTAR J163433$-$473841}, are presented in separate papers \citep{kin14,tom14}.}
\label{tab_src}
\end{deluxetable*}

\subsubsection{Systematic offset of detected sources}
\label{sec_off}

We ran \texttt{wavdetect} on individual event lists, and on the mosaic images, in order to create lists of detected sources in each energy band. In all cases, we assumed: a point-spread function (PSF) with a constant full-width at half-maximum (FWHM) of 18$^{\prime\prime}$ \citep{har13}; scale sizes of 1, 2, 4, and 8 pixels; and a threshold of $10^{-5}$. This threshold implies around 1 spurious source per observation. For the mosaics, we used the cleaned (non ghost-ray removed) images assuming a background map that mimics the observed low-frequency (i.e., large scale) ghost ray patterns with wavelet scales with characteristic sizes of 8--32\,pixels \citep{sle94,sta94,vik97}. Each pixel is 2$^{\prime\prime}$ wide, so the wavelet scales are $16^{\prime\prime}$--$64^{\prime\prime}$, i.e., larger than the high-frequency scales expected for point sources. The lack of high-frequency scales in the background map leads to a poor modeling of the sharp edges and dark dips of the ghost ray pattern. This results in a large number of source detections that align with artifacts in the image, and we conclude that they are likely spurious.

We visually inspected the event lists and the mosaic images (in each band) searching for \nustar-detected sources that were coincident with \chan\ sources (see \S\,\ref{sec_chan}). There are 3 \nustar-detected sources that have probable \chan\ counterparts. The \nustar-derived positions show a systematic offset (i.e., with a similar direction and magnitude) with respect to the \chan\ positions. In physical coordinates, this offset is $+1.98$\,pixels (3\farcs9) and $+4.75$\,pixels (9\farcs5) in the $x$ and $y$ directions, respectively, found by averaging the offsets of the \chan\ sources. This is consistent with the expected performance from \nustar\ \citep{har13}. Therefore, we registered the mosaic images to the \chan\ sources by subtracting these offset values from the reference pixel. We reran \texttt{wavdetect} to determine a final source position, positional uncertainty (quoted at 90\% confidence), and detection significance in the 3--79\,keV band. These values are reported in Table\,\ref{tab_src}.

\subsubsection{Spectral and timing analyses}

Source spectra and light curves were extracted from the cleaned event lists of each module using a 30$^{\prime\prime}$-radius circle centered on the \chan\ position while the background count rates were taken from a 90$^{\prime\prime}$-radius circle on the same detector chip: away from the source extraction region, but with a similar background pattern. The effects of vignetting on exposure were accounted for in the response matrices and spectra. The spectra were fit in \texttt{Xspec} \citep{arn96} assuming \citet{wil00} abundances and photo-ionization cross-sections of \citet{ver96}.

While this extraction radius covers roughly 40\% of the enclosed energy, the \nustar\ PSF has a relatively narrow peak (18$^{\prime\prime}$ FWHM) superimposed on broad wings, which means source extraction radii wider than this (at the off-axis angles considered here) have the undesired result of adding more background relative to the gain in source counts. Results of the spectral analysis showed that the sources emitted few counts above $\gtrsim$20\,keV, and so the energy band used for \nustar\ timing and spectral analyses was restricted to 3--24\,keV. All \nustar\ source spectra were binned to contain at least 20 net (i.e., background-subtracted) source counts and a minimum significance of 2$\sigma$.

\subsection{\emph{Chandra} Data}
\label{sec_chan}

In 2011, \chan\ observed a $\sim2.0^{\circ}\times0.8^{\circ}$ section of the Norma Arm, a subset of which is the $\sim 0.4^{\circ}\times0.4^{\circ}$ \nustar\ survey region described in this paper. Of the $\sim$1100 X-ray sources detected by Fornasini et al. (2014, subm.), we excluded all objects outside the 0.2\,deg$^{2}$ \nustar\ survey region, and then rejected those whose ratio between net source counts in the 2--10\,keV and 0.5--2\,keV energy bands was less than 0.8. This yields a catalog of 22 relatively hard sources that are suitable low-energy X-ray counterpart candidates to sources detected at higher energies with \nustar. %

Observations used in this study are ObsID \dataset[ADS/Sa.CXO#obs/12532]{12532} and ObsID \dataset[ADS/Sa.CXO#obs/12532]{12533}. Reprocessing and reduction of this data relied on CIAO v.4.5. Spectra were extracted from each event list in the 0.3--10\,keV band for a source region centered on the \chan\ position (a circle of radius $=10^{\prime\prime}$), and for a source-free background region (a rectangle with dimensions: $200^{\prime\prime}\times100^{\prime\prime}$) on the same detector chip. Spectral data were grouped to contain a minimum of 20 source$+$background counts per bin.

\begin{figure*}[!t]
\begin{center}
\includegraphics[width=0.7\textwidth]{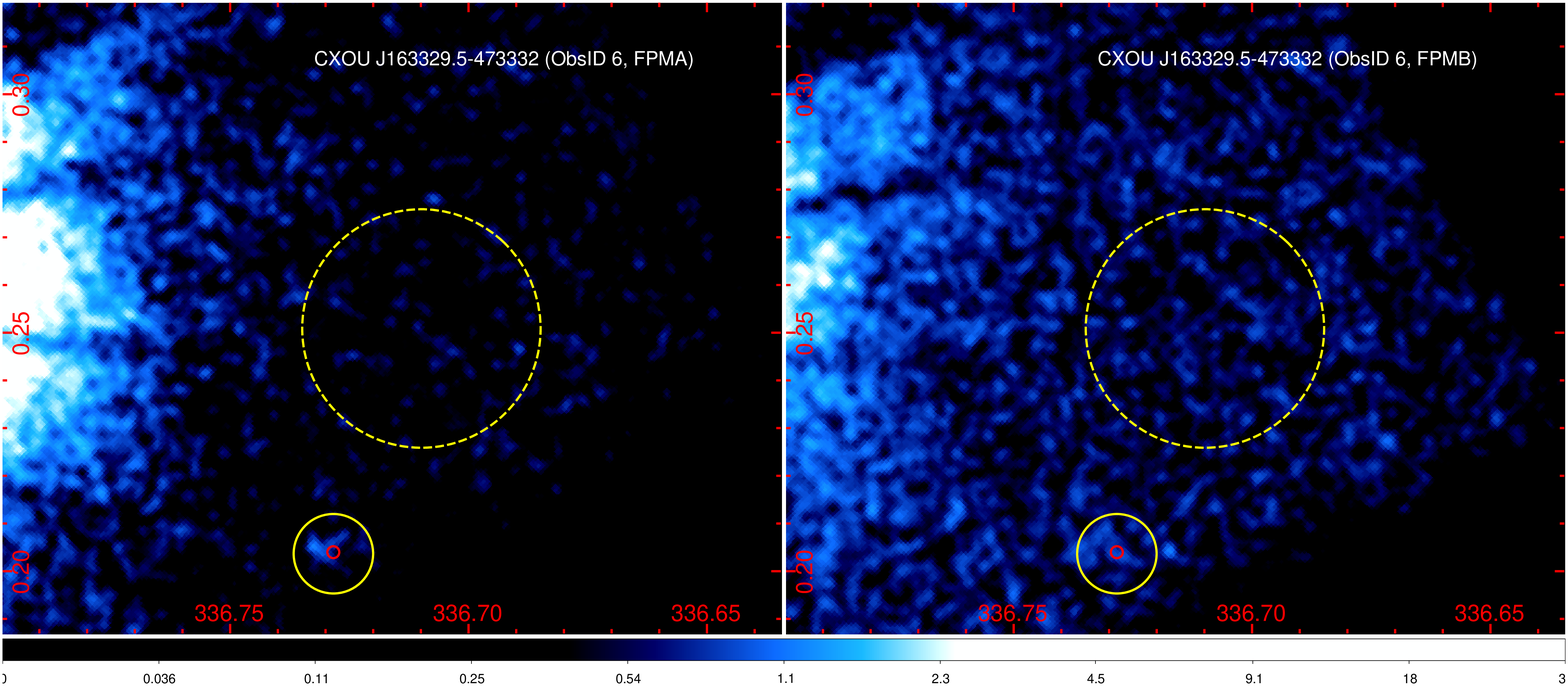}
\vspace{1mm}\vspace{0mm}
\includegraphics[width=0.7\textwidth]{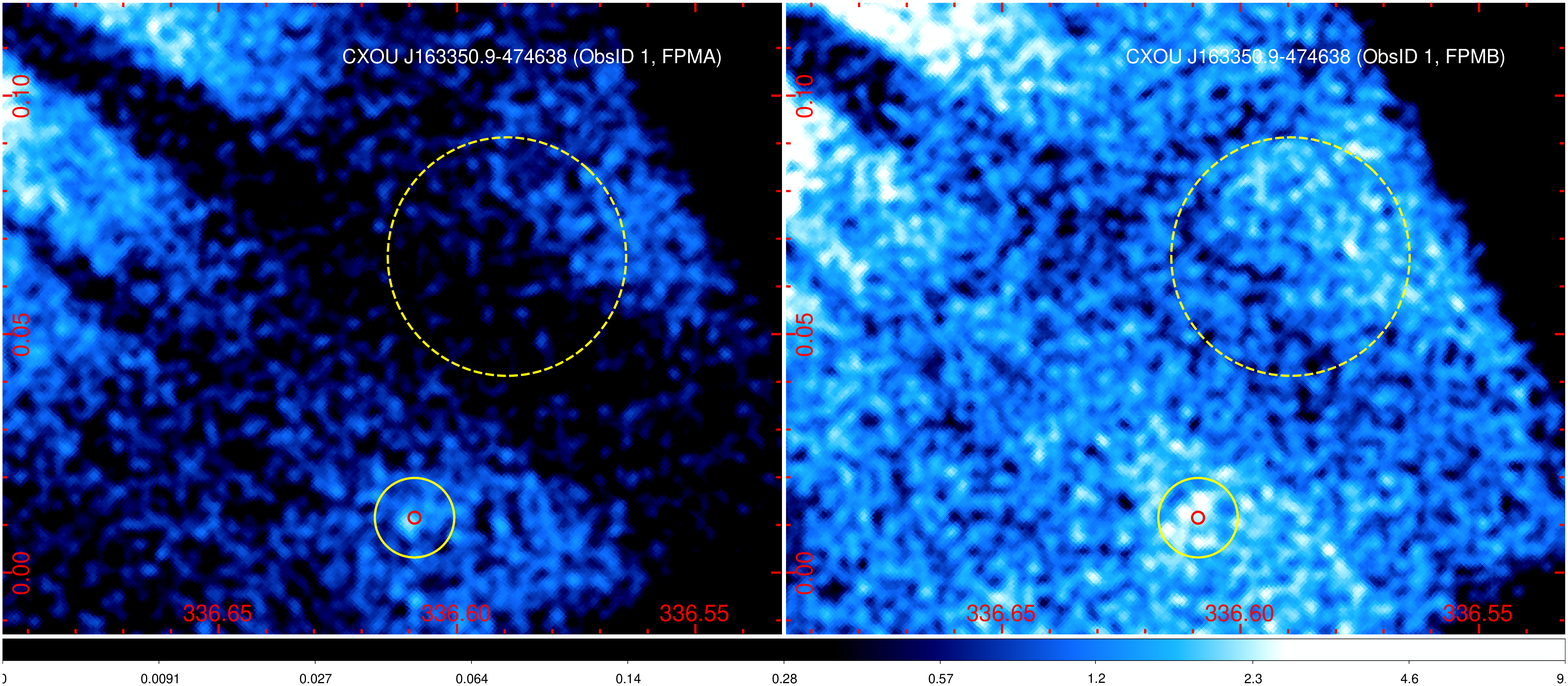}
\vspace{1mm}\vspace{0mm}
\includegraphics[width=0.7\textwidth]{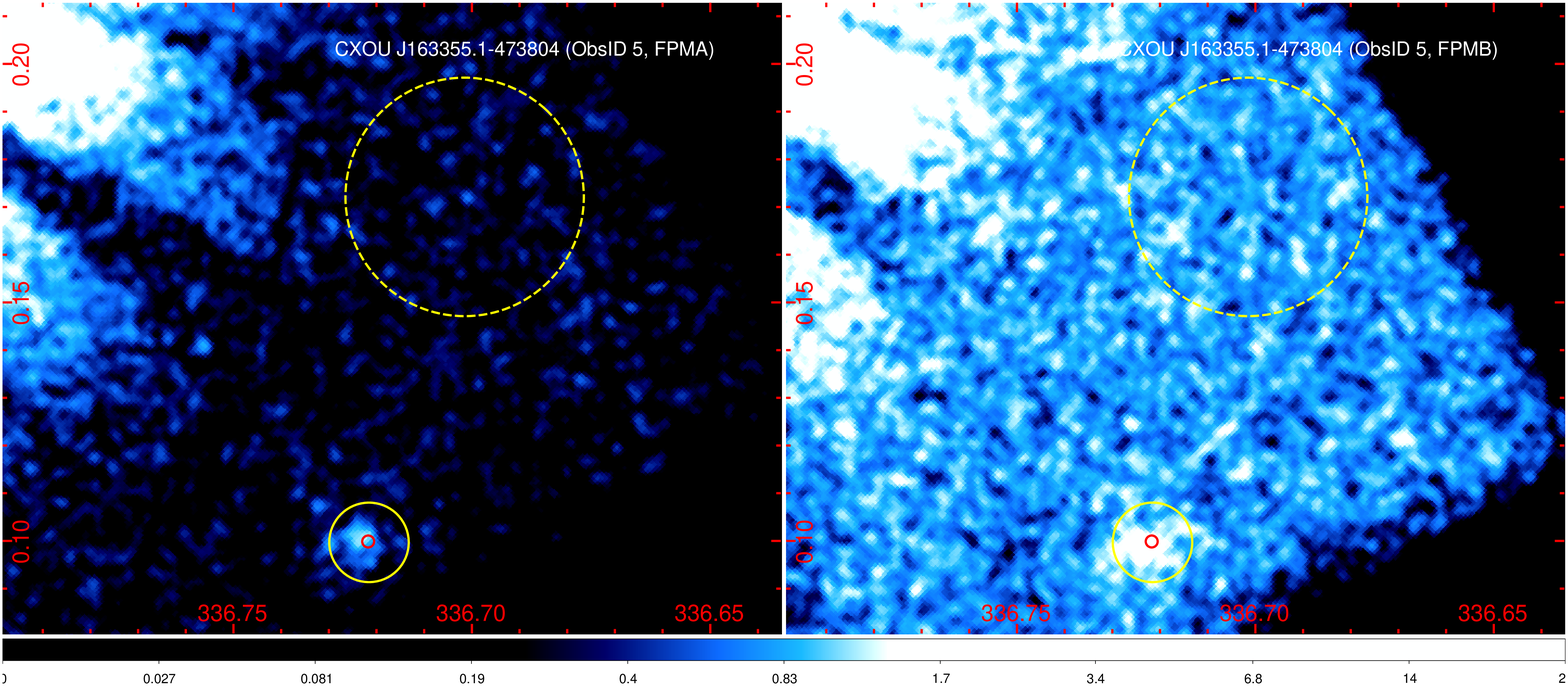}

\end{center}
\caption{\scriptsize{\nustar\ images of the sources discussed in this work in Galactic coordinates. The top, middle, and bottom rows present cleaned event lists in the 3--79-keV band from, respectively, CXOU J163329.5$-$473332 (ObsID 6), CXOU J163350.9$-$474638 (ObsID 1), and CXOU J163355.1$-$473804 (ObsID 5). The left panels show FPMA while the right panels show FPMB with the same logarithmic scaling. The images were smoothed with a Gaussian kernel of $\sigma = 6^{\prime\prime}$. The small circles indicate the \chan\ positions, the medium circles are the 30$^{\prime\prime}$ source-extraction regions, and the large dashed circles (90$^{\prime\prime}$-radius) represent the background regions.}}
\label{fig_evt}
\end{figure*}

\section{Results}
\label{sec_res}

The flux map (counts map divided by the exposure map) of the broad-band energy range (3--79\,keV) is presented in Figure\,\ref{fig_mos}. The surveying strategy, which tiled the pointings so that they contained significant overlap in their observed fields, as well as the redundancy of having two detector modules whose FOVs are slightly shifted, leads to a mosaic image that is practically free of gaps, despite the exclusion of a large fraction of pixels with stray light and ghost rays ($\sim$10\%--50\% of the pixels in each module). The photon-free region (black wedge) at the upper-left or northeast of \object{4U 1630$-$47} is due to the exclusion of pixels with ghost rays with no redundant observations that can compensate for the lack of exposure. The median exposure time is 24\,ks with the deepest regions having 96\,ks of exposure.

Although the effects of stray light and ghost rays have been minimized, the background level remains high and inhomogeneous throughout the image. The exclusion of contaminated pixels leads to artifacts that are visible as bright arcs concentric around \object{4U 1630$-$47}. Increasing the size of the exclusion region leads to exposure gaps in the mosaic. Bright fringes that appear along the right edge of the mosaic image are due to secondary ghost rays from \object{4U 1630$-$47}. The contaminated pixels are situated in a ``halo'' whose inner radius is $\gtrsim0.3^{\circ}$ from \object{4U 1630$-$47}. We did not attempt to correct for this \emph{a posteriori} due to insufficient exposure redundancy in the affected regions.

The objects in the survey region that are most easily perceptible are \object{4U 1630$-$47} and \object{NuSTAR J163433$-$473841}. Their properties are discussed in separate papers, but highlights include: the discovery of reflection from the inner accretion disk of \object{4U 1630$-$47} yielding a black hole spin of $a =$ 0.985(3), and an iron absorption feature at 7.03(3)\,keV suggesting a magnetically-driven disk wind \citep{kin14}; and the discovery of a  hard X-ray source (\object{NuSTAR J163433$-$473841}) which underwent a 1-day long X-ray flare serendipitously during our \nustar\ survey, but was never seen in any wavelength before or since those observations. This suggests that \object{NuSTAR J163433$-$473841} is a new fast X-ray transient that could be a magnetar or an active stellar binary \citep{tom14}. 

In addition to these objects, there are 3 significantly detected hard X-ray sources whose positions are compatible with sources seen at lower energies by \chan: CXOU J163329.5$-$473332, CXOU J163350.9$-$474638, and CXOU J163355.1$-$473804 (Fig.\,\ref{fig_mos}). Their basic properties are listed in Table\,\ref{tab_src}. Uncertainties are quoted at 90\% confidence, unless noted otherwise.

\begin{figure}[!t]
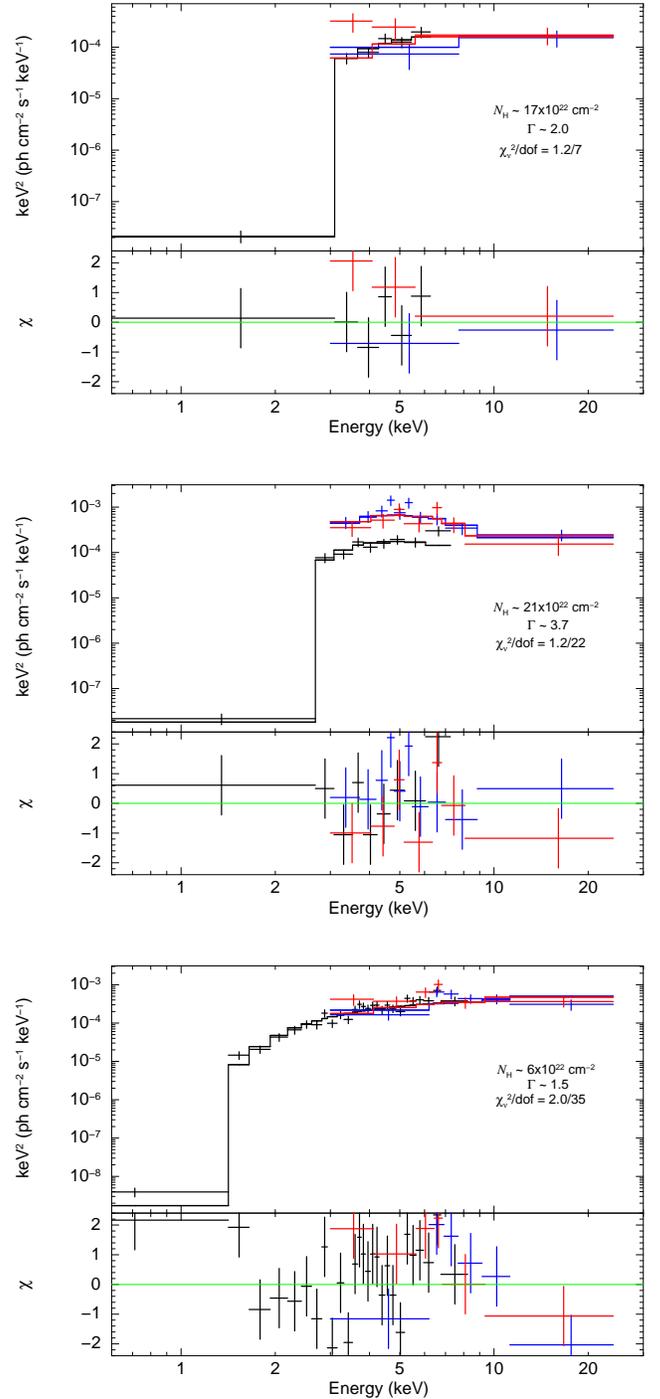

\begin{center}
\includegraphics[width=0.32\textwidth,angle=-90]{163329_spec.ps}
\vspace{3mm}\vspace{0mm}
\includegraphics[width=0.32\textwidth,angle=-90]{163350_spec.ps}
\vspace{3mm}\vspace{0mm}
\includegraphics[width=0.32\textwidth,angle=-90]{163355_spec.ps}
\end{center}
\caption{Background-subtracted spectra ($\nu F_{\nu}$) collected with \chan\ (black), \nustar-FPMA (blue), and \nustar-FPMB (red) for CXOU J163329.5$-$473332 (\emph{top}), CXOU J163350.9$-$474638 (\emph{middle}), and CXOU J163355.1$-$473804 (\emph{bottom}). Spectral bins for \chan\ contain a minimum of 20 source$+$background counts, while those of \nustar\ have at least 20 net source counts and a minimum significance of $2\sigma$. Error bars denote 90\%-confidence limits. The lower panels show residuals from absorbed power laws fit to the joint \chan-\nustar\ data. The derived spectral parameters are listed in Table\,\ref{tab_spec}.}
\label{fig_spec}
\end{figure}

\subsection{CXOU J163329.5$-$473332}

\nustar\ detects a source at the 8.3-$\sigma$ level whose position (Table\,\ref{tab_src}) is 1\farcs9 away from, and compatible with, that of \object{CXOU J163329.5$-$473332}. The source appears in \nustar\ ObsID 5 and ObsID 6. However, it falls in the chip gap and among the ghost rays during ObsID 5, and so only data from ObsID 6 was used for spectral and timing analyses (Fig.\,\ref{fig_evt}).

\chan\ spectral data from ObsID \dataset[ADS/Sa.CXO#obs/12533]{12533} were fit with an absorbed power law yields \nh\ $=(12_{-9}^{+14})\times10^{22}$\,\cmsq\ and $\Gamma = 1.2_{-1.8}^{+2.2}$ ($\chi_{\nu}^{2}/\mathrm{dof} = 0.6/3$). There were 125$\pm$12 net source counts in 0.3--10\,keV, distributed as 14$\pm$4\,cts (0.3--3\,keV) and 111$\pm$11\,cts (3--10\,keV). Using \citet{cas79} statistics and \citet{pea00} $\chi^{2}$ test statistics on unbinned \chan\ data give consistent results.

An absorbed power law was then fit to the \nustar\ data only. With \nh\ fixed to the best-fit value from \chan, we measure a photon index $\Gamma = 2.6_{-1.2}^{+1.3}$ that is consistent with the one from \chan. The source emitted 120$\pm$22 net counts in the \nustar\ energy band (3--24\,keV), and nearly all of them (105$\pm$20) were recorded below 10\,keV.

\begin{figure}[!t]
\begin{center}
\includegraphics[width=0.38\textwidth]{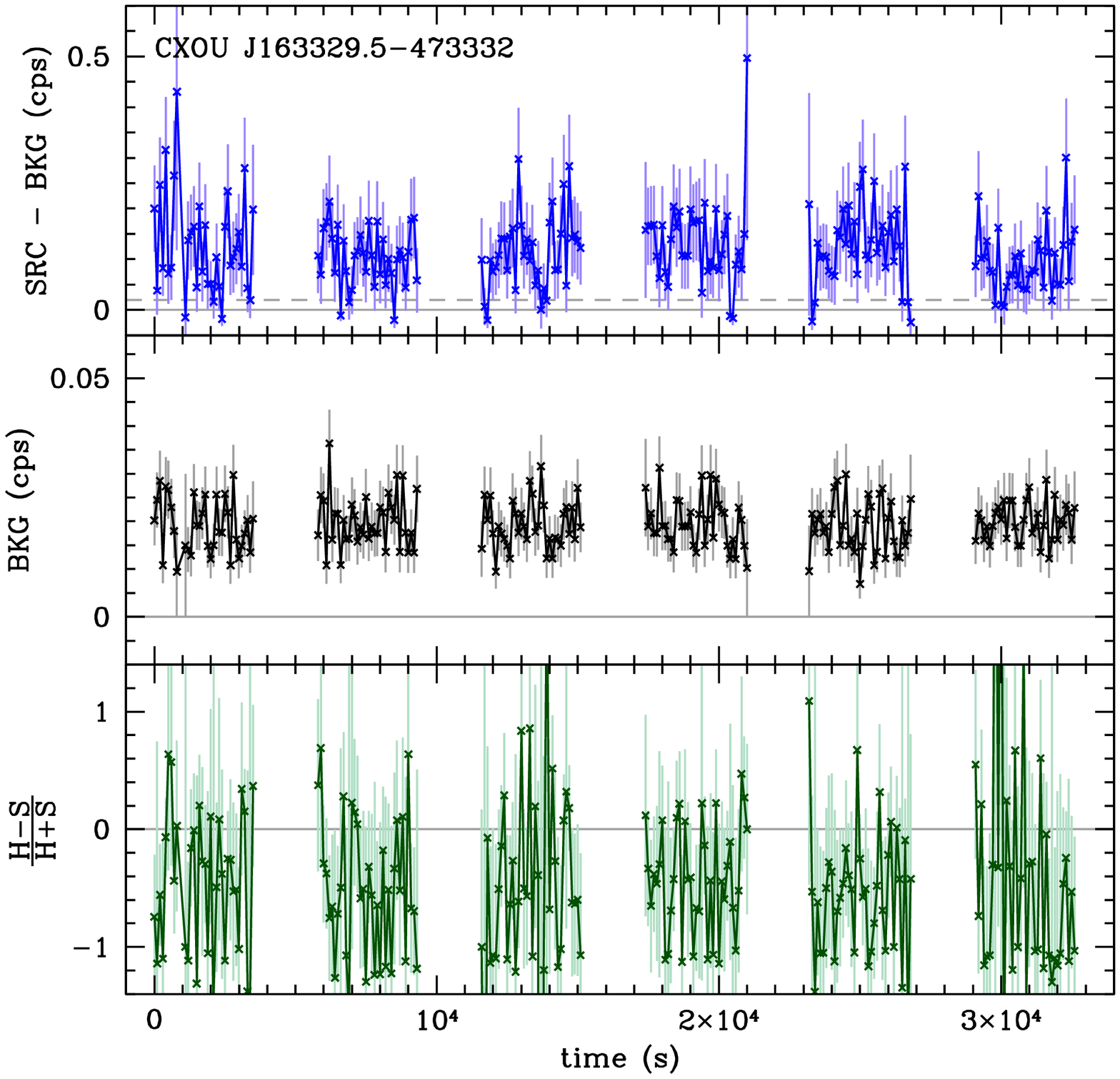}
\vspace{0mm}\vspace{0mm}
\includegraphics[width=0.38\textwidth]{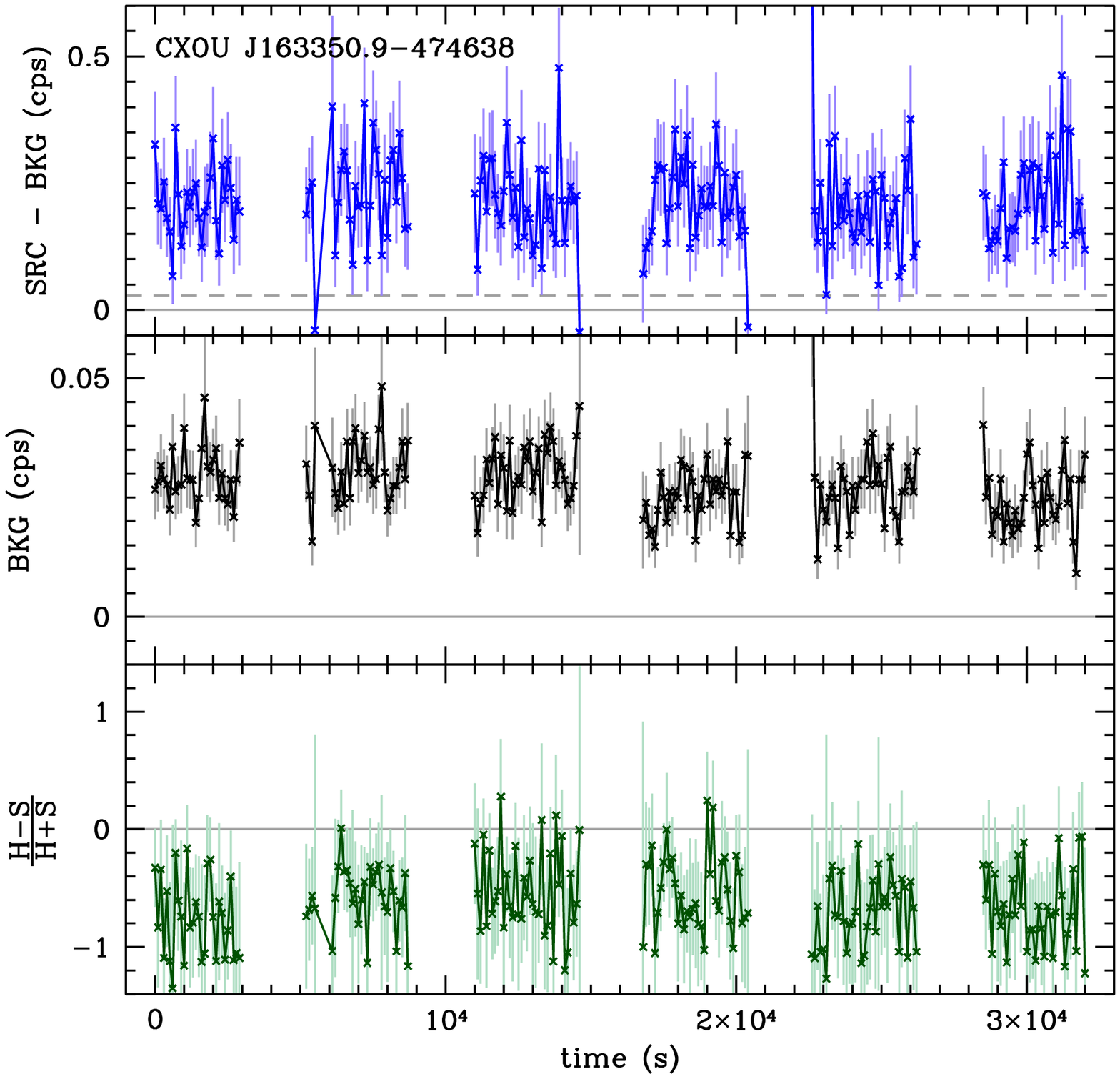}
\vspace{0mm}\vspace{0mm}
\includegraphics[width=0.38\textwidth]{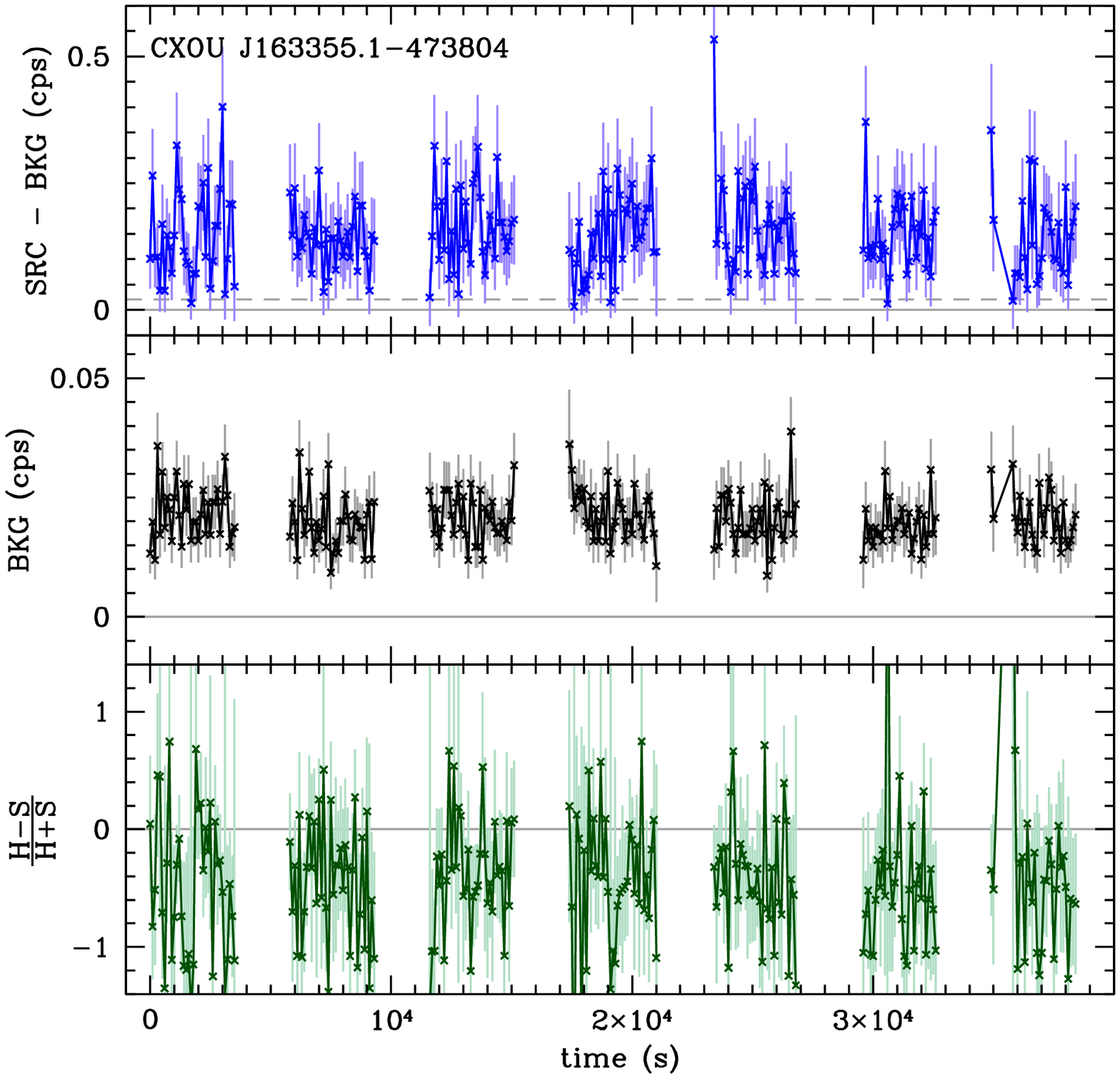}
\end{center}
\caption{Source and background light curves (3--24\,keV) for CXOU J163329.5$-$473332 (\emph{top}), CXOU J163350.9$-$474638 (\emph{middle}), and CXOU J163355.1$-$473804 (\emph{bottom}). The source light curve combines count rates from FPMA and FPMB that are then background-subtracted. The background count rate has been scaled to the size of the source region. The average background rate is shown as a dashed line in the top panel. The hardness ratio is defined as $\frac{H - S}{H+S}$ where $S$ and $H$ represent count rates in 3--8\,keV and 8--24\,keV, respectively. Each bin is 100\,s.}
\label{fig_lc}
\end{figure}

We fit the combined spectra from \chan\ and \nustar\ using a cross-instrumental constant fixed at unity for the \chan\ data, and allowed to vary for the \nustar\ data. The best-fitting model parameters for the power-law model are \nh\ $=(17_{-7}^{+10})\times10^{22}$\,\cmsq\ and $\Gamma = 2.0_{-1.1}^{+1.2}$ (Fig.\,\ref{fig_spec} and Table\,\ref{tab_spec}). A fit of equivalent quality ($\chi_{\nu}^{2}/\mathrm{dof} = 1.3/7$) is obtained with an absorbed blackbody model of temperature $kT = 2.0_{-0.6}^{+1.1}$\,keV, and a lower column density \nh\ $=(7_{-3}^{+6})\times10^{22}$\,\cmsq. The instrumental constant ($0.9_{-0.5}^{+0.7}$ for the power law, $0.9_{-0.5}^{+0.6}$ for the blackbody), which is consistent with pre-flight expectations for \nustar\ \citep{har13} and with joint \chan-\nustar\ spectral fits of other Galactic objects \citep[e.g.,][]{got14}, indicates little variability in source flux between observations taken nearly 2 years apart. Fitting the joint spectral data with a bremsstrahlung model leads to an unconstrained plasma temperature ($\approx12$\,keV). 

Figure\,\ref{fig_lc} presents the light curve (3--24\,keV, 100-s resolution) of \object{CXOU J163329.5$-$473332}. We searched for periods in the range of 0.004\,s (i.e., twice the time resolution of the light curve data used in this fine timing analysis) to 18959\,s (i.e., the observation duration), and we did not detect a significant pulsation signal in the soft (3--8\,keV), hard (8--24\,keV), or broad energy band (3--24\,keV). 

The field of \object{CXOU J163329.5$-$473332} was observed by \xmm\ and this object was detected as part of the \xmm\ Serendipitous Survey Catalog \citep{wat09}, although it appears faint and far off-axis ($\gtrsim$10 arcmin). We analyzed observation ID 0654190201 (rev. 2051) which was taken in 2011 February, with a total exposure time of 22\,ks. The parameters from the \xmm\ spectral fit of this source, i.e., the observed flux, column density, photon index, and blackbody temperature, are consistent with those derived from fits to the \chan, \nustar, and combined \chan-\nustar\ spectra.

We observed the near-infrared counterpart to \object{CXOU J163329.5$-$473332} with the NEWFIRM telescope and its magnitudes are $J =$ 15.29$\pm$0.07\,mag, $H =$ 11.92$\pm$0.10\,mag, and $K_{s} =$ 10.13$\pm$0.06\,mag \citep{rah14}. The infrared spectrum displays strong CO lines in absorption (at 16198\,$\AA$ and 22957\,$\AA$), a number of weak emission lines, and no Br-$\gamma$ line. This spectrum is typical of an early MIII-type star feeding a small accretion disk \citep{rah14}.

\begin{deluxetable*}{ l c c c c c c c c c c c  }
\tablewidth{0pt}
\tabletypesize{\scriptsize}
\tablecaption{Parameters from absorbed power law (PL), blackbody (BB), and bremsstrahlung (FF) models fit to the joint \emph{Chandra} and \emph{NuSTAR} spectral data for sources in the Norma Arm survey.}
\tablehead{
\colhead{name}  & \colhead{model} 	& \colhead{$C$ \tablenotemark{a}} & \colhead{\nh\ \tablenotemark{b}}  & \colhead{$\Gamma$  or $kT$ \tablenotemark{c}} 
							& \colhead{norm.\tablenotemark{d}} & \colhead{$\chi_{\nu}^{2}/$dof \tablenotemark{e}} 
							& \colhead{$S$ \tablenotemark{f}} & \colhead{$H$ \tablenotemark{g}} & \colhead{HR \tablenotemark{h}} 
							& \colhead{obs. flux \tablenotemark{i}} & \colhead{unabs. flux \tablenotemark{j}}  }
\startdata	

\noalign{\smallskip}

\smallskip
CXOU J163329.5$-$473332		& PL & $0.9_{-0.5}^{+0.7}$	& $17_{-7}^{+10}$		& $2.0_{-1.2}^{+1.1}$ 
							& 2.1 					& 1.2/7 		
							& 105$\pm$20 				& 15$\pm$8 			& $-$0.7$\pm$0.3 	\\

\smallskip
							
							& BB & $0.9_{-0.5}^{+0.6}$	& $7_{-3}^{+6}$			& $2.0_{-0.6}^{+1.1}$ 
							& 0.06 					& 1.3/7 		
							&   		&  				& 	
							& 10.9$\pm$2.7 			& 13.2$\pm$3.6  \\

\smallskip
							
							& FF & $0.9_{-0.5}^{+0.7}$	 & $15_{-5}^{+7}$			& $\approx 12^{\star}$ 
							& 1.3 					& 1.2/7 		
							&   		&  				& 	
							& 12.0$\pm$2.1$^{\star}$		& 21.2$\pm$5.9$^{\star}$  \\

\smallskip
CXOU J163350.9$-$474638		& PL & $3.8_{-0.7}^{+0.9}$	& $21_{-4}^{+6}$		& 3.7$\pm$0.5 
							& 41.3					& 1.1/22 		
							& 375$\pm$28 				& 25$\pm$10			& $-0.9_{-0.1}^{+0.2}$	
							& 29.1$\pm$9.3 			& 362$\pm$238 \\

\smallskip							
							& BB & $3.6_{-0.7}^{+0.9}$	& $9_{-2}^{+3}$			& $1.2_{-0.1}^{+0.2}$ 
							& 0.06 					& 1.3/22 		
							&   		&  				& 	
							& 25.8$\pm$2.2 			& 39.6$\pm$10.9 \\

\smallskip							
							& FF & $3.7_{-0.7}^{+0.9}$	& $15_{-3}^{+4}$			& $3.3_{-0.7}^{+1.0}$ 
							& 4.0 					& 1.1/22 		
							&   		&  				& 	
							& 27.6$\pm$5.3			& 77.2$\pm$31.3 \\

\smallskip

CXOU J163355.1$-$473804 \tablenotemark{$\dagger$}		& PL & 1.0$\pm$0.3			& 6$\pm$1			& 1.5$\pm$0.3 
							& 1.5						& 2.0/35 		
							& 256$\pm$26				& 52$\pm$12			& $-$0.7$\pm$0.2
							& 32.8$\pm$3.2 			& 43.2$\pm$7.1 \\
							
\smallskip
							& BB & 1.0$\pm$0.2			& $1.2_{-0.4}^{+0.5}$			& $2.1_{-0.2}^{+0.3}$ 
							& 0.1 					& 1.3/35 		
							&   		&  				& 	
							& 28.1$\pm$3.7			& 29.1$\pm$2.8 \\

\smallskip
							& FF & 1.1$\pm$0.3			& $5.2_{-0.9}^{+1.1}$			& $21_{-9}^{+32}$ 
							& 1.9 					& 1.8/35 		
							&   		&  				& 	
							& 32.7$\pm$10.3			& 41.0$\pm$18.6
									
\enddata

\tablenotetext{a}{Instrumental constant fixed to 1 for the \chan\ data and allowed to vary for the \nustar\ data. }
\tablenotetext{b}{Column density in units of $10^{22}$\,\cmsq. }
\tablenotetext{c}{Photon index of the power law (PL) model, or plasma temperature (in keV) for the blackbody (BB) and bremsstrahlung (FF) models.}
\tablenotetext{d}{Model normalization ($\times10^{-4}$).}
\tablenotetext{e}{Reduced $\chi^{2}$ over degrees of freedom (dof).}
\tablenotetext{f}{Net source counts from both \nustar\ modules combined in the soft ($S$) band: 3--10\,keV.}
\tablenotetext{g}{Net source counts from both \nustar\ modules combined in the hard ($H$) band: 10--24\,keV.}
\tablenotetext{h}{Hardness ratio defined as $(H-S)/(H+S)$.}
\tablenotetext{i}{Observed flux (i.e., not corrected for absorption) in units of $10^{-13}$\,\ergcms\ in the 0.3--24\,keV band.}
\tablenotetext{j}{Absorption-corrected flux in units of $10^{-13}$\,\ergcms\ in the 0.3--24\,keV band.}
\tablenotetext{$\star$}{The fluxes are derived by fixing the plasma temperature to 12\,keV.}
\tablenotetext{$\dagger$}{The best fitting model for \object{CXOU J163355.1$-$473804} requires a Gaussian component at 6.7\,keV.}
\label{tab_spec}
\end{deluxetable*}

\subsection{CXOU J163350.9$-$474638}

In ObsID 1 (Fig.\,\ref{fig_evt}), \nustar\ detects a source at a significance of 15.0$\sigma$ (Table\,\ref{tab_src}) whose position is 4\farcs1 from, and compatible with, the \chan\ position of \object{CXOU J163350.9$-$474638}. 

The \chan\ spectral data (ObsID \dataset[ADS/Sa.CXO#obs/12532]{12532}) were fit with an absorbed power law to give \nh\ $= (13_{-5}^{+8})\times10^{22}$\,\cmsq\ and $\Gamma = 2.0_{-1.1}^{+1.3}$ ($\chi_{\nu}^{2}/\mathrm{dof} = 0.8/6$). A thermal blackbody model ($kT = 1.4_{-0.4}^{+0.7}$\,keV) fit to the data yields a similar column (\nh\ $=(8_{-3}^{+5})\times10^{22}$\,\cmsq) and fit quality ($\chi_{\nu}^{2}/\mathrm{dof} = 1.1/6$). There are 190$\pm$14 net source counts in the 0.3--10\,keV range, with 30$\pm$6 counts having energies below 3\,keV, and the rest (159$\pm$13) are above 3\,keV.

We then fit an absorbed power law to the \nustar\ data alone while fixing \nh\ to the best-fit value from \chan. The fit quality is decent ($\chi_{\nu}^{2}/\mathrm{dof} = 1.2/14$) and the photon index ($\Gamma =$ 3.3$\pm$0.3) is consistent with the value measured with \chan. An absorbed blackbody provides an acceptable fit ($\chi_{\nu}^{2}/\mathrm{dof} = 1.4/22$) with a temperature $kT =$ 1.1$\pm$0.1\,keV, similar to that measured with \chan. The source emitted 400$\pm$30 net counts in 3--24\,keV, with most of them (375$\pm$28) below 10\,keV.

The spectra from \chan\ and \nustar\ were jointly fit with an absorbed power law yielding \nh\ $=(21_{-4}^{+6})\times10^{22}$\,\cmsq\ and $\Gamma =$ 3.7$\pm$0.5 (Fig.\,\ref{fig_spec} and Table\,\ref{tab_spec}). Although the fit quality is good ($\chi_{\nu}^{2}/\mathrm{dof} = 1.1/22$), the cross-instrumental constant is $3.8_{-0.7}^{+0.9}$ which indicates significant variability on year-long timescales. 

Adding an exponential cutoff constrains the break energy ($E_{\mathrm{cut}} \le 13$\,keV). However, this component is not required by the data since it returns a similar $\chi_{\nu}^{2}$ with 2 less dof. The measured \nh\ is larger in the joint fit than in the \chan\ data alone, and fixing the column density to the \chan\ value leads to a poorer fit ($\chi_{\nu}^{2}/\mathrm{dof} = 1.7/23$). 

Thermal models also provide good fits to the data. A blackbody model ($\chi_{\nu}^{2}/\mathrm{dof} = 1.3/22$) gives a lower column density than for the power law (\nh\ $=(9_{-2}^{+3})\times10^{22}$\,\cmsq), and has a temperature of $kT = 1.2_{-0.1}^{+0.2}$\,keV. A bremsstrahlung model ($\chi_{\nu}^{2}/\mathrm{dof} = 1.1/22$) has an absorbing column consistent with the power law model (\nh\ $=(15_{-3}^{+4})\times10^{22}$\,\cmsq), and a plasma temperature of $kT = 3.3_{-0.7}^{+1.0}$\,keV (a value that is not constrained with the \chan\ data alone). 

The 3--24-keV light curve binned at 100\,s is presented in Fig.\,\ref{fig_lc}, and it shows \object{CXOU J163350.9$-$474638} to be a relatively soft source that displays low variability on short timescales. The background is mostly due to \object{4U 1630$-$47} whose ghost-ray halo covers the extraction regions used to produce the light curves. The apparent decrease in background counts is not significant. There are no periodicities detected in the range of 0.004\,s to 18407\,s in any energy range. An upper limit of $\sim$30\% (at 90\% confidence) is derived for the fractional r.m.s. expected for a periodic signal.

\subsection{CXOU J163355.1$-$473804}

This source appears in two \chan\ observations (ObsID \dataset[ADS/Sa.CXO#obs/12532]{12532} and ObsID \dataset[ADS/Sa.CXO#obs/12533]{12533}); the spectral data from these observations were summed to give 546$\pm$24 net source counts (0.3--10\,keV), divided into 168$\pm$13 and 377$\pm$20 net counts in the 0.3--3\,keV and 3--10\,keV bands, respectively. A power law model fit to the binned spectral data provides an adequate fit ($\chi_{\nu}^{2}/\mathrm{dof} = 1.3/24$) with \nh\ $= (3\pm1)\times10^{22}$\,\cmsq\ and a flat photon index: $\Gamma = 0.7_{-0.3}^{+0.4}$. A blackbody of temperature $kT = 1.9_{-0.3}^{+0.4}$\,keV improves the fit slightly ($\chi_{\nu}^{2}/\mathrm{dof} = 1.2/24$).

The likely hard X-ray counterpart to \object{CXOU J163355.1$-$473804} is detected at the 8.7-$\sigma$ level (3--79\,keV) in the \nustar\ mosaic image. Ghost-ray photons contaminate the region around the source in ObsID 4, and so spectral and timing analysis relied only on data from ObsID 5. The 321$\pm$28 net source counts (3--24\,keV) were distributed as 256$\pm$26 net counts in 3--10\,keV, and 52$\pm$12 in 10--24\,keV. 

We fit the \nustar\ data with power law and blackbody models holding \nh\ fixed to the best-fit value from \chan, and derived a steeper photon index ($\Gamma =$ 1.9$\pm$0.3) or a plasma temperature consistent with that of \chan\ ($kT =$ 2.2$\pm$0.3\,keV), with both models giving poor fits ($\chi_{\nu}^{2}/\mathrm{dof} = 2.3/9$ and $\chi_{\nu}^{2}/\mathrm{dof} = 1.8/9$, respectively). 

Jointly fitting the \chan\ and \nustar\ data gives a poor fit ($\chi_{\nu}^{2}/\mathrm{dof} = 2.0/35$) when using only an absorbed power law: \nh\ $= (6\pm1)\times10^{22}$\,\cmsq\ and $\Gamma =$ 1.5$\pm$0.3 (Fig.\,\ref{fig_spec} and Table\,\ref{tab_spec}). The fit quality improves ($\chi_{\nu}^{2}/\mathrm{dof} = 1.3/35$) with a blackbody model having $kT = 2.1_{-0.2}^{+0.3}$\,keV and a lower column density (\nh\ $= (1.2_{-0.4}^{+0.5})\times10^{22}$\,\cmsq), or with a power law and exponential cutoff ($\chi_{\nu}^{2}/\mathrm{dof} = 1.4/30$) where \nh\ $= (3\pm1)\times10^{22}$\,\cmsq, $\Gamma = 0.6\pm0.4$, and the cutoff energy is $5_{-1}^{+3}$\,keV. In both cases, the instrumental constant is 1.1$\pm$0.2 suggesting little variability over yearlong timescales. 

Residuals remain around 6.7\,keV where emission from the fluorescence of ionized iron is expected. Indeed, the best spectral fits are obtained when a Gaussian component ($\sigma = 0$) is added to either the cutoff power law or the blackbody model. In order to analyze this line, we rebinned the \nustar\ spectra to have at least 20 source$+$ background counts and a minimum significance of 2$\sigma$. For the power law ($\chi_{\nu}^{2}/\mathrm{dof} = 1.1/48$), \nh\ $= (2\pm1)\times10^{22}$\,\cmsq, with $\Gamma = 0.0_{-1.0}^{+0.6}$ and an exponential cutoff at $4_{-1}^{+3}$\,keV. The line centroid is $6.72_{-0.08}^{+0.04}$\,keV with an equivalent width of $\sim$500\,eV (unconstrained). 

For the blackbody model ($\chi_{\nu}^{2}/\mathrm{dof} = 1.2/50$), the line centroid is $6.7_{-0.2}^{+0.1}$\,keV with an equivalent width of $414_{-312}^{+370}$\,eV. The column density and blackbody temperature are \nh\ $= (1.2_{-0.5}^{+0.6})\times10^{22}$\,\cmsq\ and $kT = 2.0_{-0.2}^{+0.3}$\,keV, respectively. The radius of the emitting region implied by the blackbody model is very small (0.03--0.16\,km assuming source distances of 2--10\,kpc). Either the source is very distant ($\ge$20\,kpc) or the blackbody is not the right model.  

We replaced the blackbody continuum with a bremsstrahlung ($\chi_{\nu}^{2}/\mathrm{dof} = 1.3/50$) and obtained \nh\ $= (5\pm1)\times10^{22}$\,\cmsq, $kT = 16_{-5}^{+14}$\,keV, a line energy of $6.74_{-0.06}^{+0.05}$\,keV, and an equivalent width of $911_{-365}^{+553}$\,eV. We also modeled the continuum with \texttt{apec} ($\chi_{\nu}^{2}/\mathrm{dof} = 1.3/51$) and the resulting iron abundance is at least 40\% greater than Solar ($N_{\mathrm{Fe}} \ge 1.4$) with a plasma temperature of $kT = 12_{-3}^{+7}$\,keV.

The \nustar\ light curve (3--24\,keV) for \object{CXOU J163355.1$-$473804} is presented in Fig.\,\ref{fig_lc}. No coherent pulsations were detected for search periods ranging from 2\,ms to $\sim$21\,ks. 

The likely infrared counterpart to \object{CXOU J163355.1$-$473804} was observed with NEWFIRM giving magnitudes of $J =$ 16.43$\pm$0.07\,mag, $H =$ 15.45$\pm$0.10\,mag, and $K_{s} =$ 14.99$\pm$0.09\,mag \citep{rah14}. With a weak CO line at 16198\,$\AA$, a strong CO line at 22957\,$\AA$, and weak Br-$\gamma$ emission, the infrared spectrum is typical of a late GIII-type star.

\section{Discussion}
\label{sec_disc}

\subsection{CXOU J163329.5$-$473332}

The \chan\ position for \object{CXOU J163329.5$-$473332} is encompassed by the 2\farcm1 uncertainty radius of an \integ-detected source named IGR J16336$-$4733 \citep{kri10} which was also detected in a short observation by \swift\ \citep{lan11}. The flux recorded by \swift-XRT (2--10\,keV) and by \nustar\ (3--10\,keV) translate to X-ray luminosities of $1.9\times10^{34} \left [ \frac{d}{\mathrm{10\,kpc}} \right ]^{2} $\,\ergs, and $7.9\times10^{33} \left [ \frac{d}{\mathrm{10\,kpc}} \right ]^{2} $\,\ergs, respectively. The available X-ray data of \object{CXOU J163329.5$-$473332} show it to be a faint, absorbed (\nh\ $\gtrsim10^{23}$\,\cmsq), and relatively hard X-ray source (the bulk of its photons are emitted in 3--10\,keV). 

Thus, \object{CXOU J163329.5$-$473332} could be a faint low-mass X-ray binary \citep[LMXB: e.g.,][]{deg09} or a cataclysmic variable \citep{kuu06} of the intermediate polar \citep[IP: e.g.,][]{pat94} variety due to the hard X-ray detection. The detection of \object{CXOU J163329.5$-$473332} out to $\sim$20\,keV with a moderately steep photon index ($2.4_{-0.8}^{+0.9}$) and low X-ray luminosity is consistent with both classifications. Another possibility is a binary system in which the compact object is a non-accreting magnetar \citep[e.g.,][]{tho96}.

\subsection{CXOU J163350.9$-$474638}

These \nustar\ observations of \object{CXOU J163350.9$-$474638} extend the source spectrum beyond 10\,keV. However, the source demonstrates significant variability in intensity (by at least a factor of 4) over the two years separating the \chan\ and \nustar\ observations, which makes it difficult to draw firm conclusions from joint-fitting of the broadband X-ray spectral energy distribution. 

Nevertheless, it is possible to compare the spectral parameters derived from single-instrument fits. The photon index is steeper in the \nustar\ data (by $\sim$50\%) compared with the value measured with \chan. This is not uniquely due to the fact that \nustar\ covers higher X-ray energies, since $\sim$90\% of the photons recorded by \nustar\ were below 10\,keV, i.e., in an energy range covered by \chan. On the other hand, thermal models also fit the data well, and the blackbody temperature ($kT =$ 1.2$\pm$0.2\,keV) and column density (\nh\ $= (9_{-4}^{+5})\times10^{22}$\,\cmsq) are in agreement for both \chan\ and \nustar\ spectra. 

There are no catalogued IR/optical objects from Vizier\footnote{http://vizier.u-strasbg.fr} or in the Vista Variables in the Via Lactae Survey \citep{min10} compatible with the \chan\ position. Thus, \object{CXOU J163350.9$-$474638} lacks a stellar counterpart which would rule out a CV or XRB located nearby, while the steep power law disfavors an AGN. Given its thermal spectrum, its long-term variability, and the absence of multi-wavelength counterparts, we conclude that \object{CXOU J163350.9$-$474638} could be a low-mass X-ray binary situated a large distance away, or perhaps an isolated, magnetized NS (i.e., a magnetar).

\subsection{CXOU J163355.1$-$473804}

Prior to the \nustar\ survey, \chan\ found \object{CXOU J163355.1$-$473804} to be a relatively bright X-ray source with a hard spectral continuum. As the brightest of the three objects in this study, this permitted us to measure the source's broadband X-ray spectrum with relatively high precision. The spectrum combining \chan\ and \nustar\ data is consistent with a cutoff power law of $\Gamma = 0.0_{-1.0}^{+0.6}$ and $E_{\mathrm{cut}}=4_{-2}^{+3}$\,keV. Thermal models such as a blackbody with $kT = 2.0_{-0.2}^{+0.3}$\,keV or a bremsstrahlung with $kT = 16_{-5}^{+14}$\,keV also describe the data well, although the implied size of the emission region is not consistent with the blackbody model. The column density required by the best-fitting models (\nh\ $\lesssim 3\times10^{22}$\,\cmsq) is lower than measured for the two other sources in the study, indicating that the source is either less intrinsically absorbed than the others, or more likely, that it is closer to us.

With \nustar, we are able to confirm the detection of an iron line that is hinted at in the \chan\ data. The line energy of 6.7\,keV suggests thermal $K\alpha$ emission from highly ionized, helium-like iron (Fe\,XXV) in the optically thin plasma around an accreting white dwarf, i.e., a  cataclysmic variable \citep[e.g.,][]{hel04,pan05,kuu06}. For example, \object{EX Hya} and \object{V405 Aur} are CVs that show a 6.7\,keV line with equivalent widths $\sim$400--900\,eV, i.e., consistent with the equivalent width measured in \object{CXOU J163355.1$-$473804} \citep{hel98}.

The identification of the infrared counterpart as a cool, GIII star supports the CV classification. Another factor favoring a CV nature for \object{CXOU J163355.1$-$473804} is the apparent lack of change in intensity or spectrum during the two years separating the \chan\ and \nustar\ surveys, with no indication from all-sky X-ray monitors that the system underwent a major outburst ($L_{X} \gtrsim 10^{36}$\,\ergs) in that time (or at any time in the past few decades). 

Its lower absorbing column compared with the other sources in the survey suggests that \object{CXOU J163355.1$-$473804} is at a distance of 2 or 3\,kpc at most, i.e., in the Crux Arm, or in the nearest arc of the Norma Arm. At an assumed distance of 3\,kpc, the absorption-corrected flux (0.3--79\,keV) of the bremsstrahlung model translates to an X-ray luminosity of $5\times10^{33}$\,\ergs. This is consistent with the persistent X-ray luminosity expected from a CV \citep[e.g.,][]{mun04,kuu06}.

\subsection{Undetected \emph{Chandra} sources}

Of the 22 hard \chan\ sources in the survey region, 3 were detected by \nustar, and they ranked first, second, and fourth in order of the number of hard X-ray ($\ge$3\,keV) counts recorded by \chan. The third brightest source in the hard \chan\ band is \object{CXOU J163358.9$-$474214}. This source was not detected in the \nustar\ event lists and mosaic images, despite the fact that it was located in a relatively ghost-ray free and stray-light free part of the image in ObsID 1. This indicates a variable nature for this object (significant variability was also observed with \chan), and we establish a 3-$\sigma$ upper limit of $7\times10^{-13}$\,\ergcms\ on the absorbed source flux in the 3--10\,keV range, i.e., higher than the average flux registered by \chan\ in a similar energy band (2--10\,keV: Fornasini et al., 2014, subm.). The \chan\ error circle for this source contains a counterpart candidate seen in the near-IR by 2MASS, and in the mid-IR by \emph{Spitzer} and \emph{WISE}. The X-ray variability and the possible association with an IR-emitting source suggest a low-mass X-ray binary or a cataclysmic variable.

All other \chan\ sources in the \nustar\ survey region had less than 35\,cts in the hard \chan\ band which means they are too faint to be detected by \nustar\ given the exposure depth of this survey. 

Sections of the Norma field have been observed by \xmm\ and source candidates found therein are listed in the \xmm\ Serendipitous Survey Catalog \citep{wat09}. Of the $\sim$150 sources in the field, 22 of these are both relatively bright (flux in the 0.2--12\,keV band $\ge 5\times10^{-13}$\,\ergcms) and hard (hardness ratio between the 2--4.5\,keV and 4.5--12\,keV bands $\ge$ 0.0). Only one of them coincides spatially with the error circle of a \nustar\ source: \object{CXOU~J163329.5$-$473332}. It is one of the hardest sources (ranked 6th hardest out of 22), but it is also among the faintest (ranked 19th in flux out of 22).

\subsection{Lessons learned from this pilot study}

Besides the analysis of X-ray sources, one of the primary goals of this pilot study is to optimize the strategy for future observations. Our experiences with this mini-survey showed us that some of our strategic choices were sound and some can be improved.

Based on the results of the \chan\ survey, we knew that the mini-survey region contained several sources that \nustar\ could detect. As was done here, observers should select regions in such a way that they encompass the largest number of hard \chan\ sources (or, when available, \xmm\ sources) that are relatively bright, but not so bright that their ghost rays and stray light contaminate adjacent observations. With the exposures available in this survey (10--100\,ks), \nustar\ was able to detect three out of four X-ray sources that had more than $\sim$100\,cts in the hard \chan\ band ($\ge$3\,keV). The non-detection of the fourth source still gives the useful result that the source is variable. While this \chan\ hard-band count rate could be used as a rule-of-thumb for a source's detectability in a typical mini-survey such as this, it is no guarantee since it does not account for X-ray sources that are variable, or that were in the soft state during the \nustar\ survey.

Another factor that led to the selection of this region was that we expected it to contain a relatively low level of stray light given the satellite's roll angle at the time the observations were performed. Even if stray light were to affect one or both of the modules, substitute coverage is available from the overlapping module and/or adjacent observation(s).

The value of exposure redundancy, not only thanks to the two modules but also by tiling observations with significant overlap ($\sim$50\% shifts), can not be overstated for eliminating or reducing imaging artifacts. This is an important factor that greatly facilitated the analysis of the faint sources in this study. Further improvements in this direction can be made by dividing up the 25\,ks exposures into two or three 10--15\,ks exposures tiled with slightly more overlap (roughly $2/3$) between adjacent pointings. While data with more overlap will take more time to analyze (i.e., the spectra from separate observations will need to be merged to obtain meaningful statistics) the tradeoff is increased exposure redundancy in case pixels need to be discarded due to ghost rays or stray light.

Observers who wish to use \nustar\ for galactic surveys can prevent or reduce the effects of stray light and ghost rays in two ways: 1) by using opportunistic observations gathered only when known transients are off or emitting at low levels according to wide-field X-ray monitors such as \maxi, \swift-BAT, and \integ-ISGRI; and 2) by increasing the exposure redundancy. While we underestimated its effects during the planning of this survey, we now know more about the brightness and extent of the ghost-ray pattern from objects such as \object{4U~1630$-$47} which will help guide the selection of future surveys.

An open question is whether \nustar\ should continue to survey ``regions'' rather than using the observing time to place the most promising targets from these regions on axis. However, it is important to note that a targeted approach might have missed the discovery of the new X-ray transient \object{NuSTAR J163433$-$473841}.

While there are technical challenges, there are also tremendous scientific benefits from surveying the Galaxy with \nustar. Understanding the disk-wind connection in \object{4U 1630$-$47}, the serendipitous discovery of \object{NuSTAR J163433$-$473841}, and insights into the faint members of the galactic X-ray population are primary among these. Surveys allow \nustar\ to offer a complete picture of the Inner Milky Way which will add to our knowledge of the content of our host galaxy and unlock new mysteries.

\section{Summary \& Conclusions}
\label{sec_conc}

An initial \nustar\ survey of the Norma Arm gave insights into the hard X-ray spectral and timing behavior of five sources, three of which are described for the first time in this paper. These three sources have unclassified soft X-ray counterparts from \chan, so the broadband 0.3--79\,keV data (including IR follow-up observations) allow us to propose their likely classifications. 

As a faint, hard X-ray source with a low-mass companion, \object{CXOU J163329.5$-$473332} is shown to be either a cataclysmic variable or a faint low-mass X-ray binary. The intensity variations on year-long timescales and the lack of a clear multi-wavelength counterpart indicate that \object{CXOU J163350.9$-$474638} could be a distant X-ray binary or possibly a magnetar. We discovered a helium-like iron line at 6.7\,keV in the \nustar\ spectrum of \object{CXOU J163355.1$-$473804}, and so it is classified as a nearby cataclysmic variable given the low mass of its IR counterpart.

With \nustar\,we are granted unprecedented views into the hard X-ray populations of our Galaxy. While \nustar\ can perform surveys, its observations can be affected by ghost rays and stray light. These effects can be diminished by planning observations to avoid bright sources located just outside the field of view, and by increasing the exposure redundancy. More \nustar\ surveys are planned for the Norma Arm and other crowded fields such as the Galactic Center, and those observations will benefit from the lessons learned during this pilot study.

\acknowledgments
The authors thank the anonymous referee whose constructive review led to significant improvements in the manuscript. This work was supported under NASA Contract No. NNG08FD60C, and made use of data from the \nustar\ mission, a project led by the California Institute of Technology, managed by the Jet Propulsion Laboratory, and funded by the National Aeronautics and Space Administration. We thank the \nustar\ Operations, Software, and Calibration teams for support with the execution and analysis of these observations. This research has made use of the \nustar\ Data Analysis Software (NuSTARDAS) jointly developed by the ASI Science Data Center (ASDC, Italy) and the California Institute of Technology. This research has made use of: data obtained from the High Energy Astrophysics Science Archive Research Center (HEASARC) provided by NASA's Goddard Space Flight Center; NASA's Astrophysics Data System Bibliographic Services; and the SIMBAD database operated at CDS, Strasbourg, France. FMF acknowledges support from the National Science Foundation Graduate Research Fellowship. FEB acknowledges support from Basal-CATA PFB-06/2007, CONICYT-Chile (FONDECYT 1141218 and "EMBIGGEN" Anillo ACT1101), and Project IC120009 "Millennium Institute of Astrophysics (MAS)" of Iniciativa Cient\'{\i}fica Milenio del Ministerio de Econom\'{\i}a, Fomento y Turismo.

\bibliographystyle{apj}
\bibliography{bod.bib}

\end{document}